\documentclass[12pt]{article}
\begin{document}
\begin{center}
{\large\bf Modified Gravitational Theory as an Alternative to Dark Energy
and Dark Matter}
\vskip 0.3 true in {\large J. W. Moffat}
\vskip 0.3 true
in {\it The Perimeter Institute for Theoretical Physics, Waterloo,
Ontario, N2J 2W9, Canada}
\vskip 0.3 true in and
\vskip 0.3 true in {\it
Department of Physics, University of Toronto, Toronto, Ontario M5S 1A7,
Canada}
\end{center}
\begin{abstract}%
The problem of explaining the acceleration of the expansion of the
universe and the observational and theoretical difficulties associated with
dark matter and dark energy are discussed. The possibility that
Einstein gravity does not correctly describe the large-scale structure of
the universe is considered and an alternative gravity theory is proposed
as a possible resolution to the problems.
\end{abstract}
\vskip 0.2 true in
e-mail: jmoffat@perimeterinstitute.ca


\section{Introduction}

The surprising observational discovery that the expansion of
the universe is accelerating~\cite{Perlmutter,Riess,Spergel} has led to an
increasing theoretical effort to understand this phenomenon. Although
interpreting the data by postulating a non-zero, positive
cosmological constant is the simplest way to understand the data, it is not
satisfactory, because it leads to the two serious problems of why
the estimates from the standard model and quantum field theory predict
preposterously large values of the cosmological constant~\cite{Weinberg},
and the coincidence of dark energy dominance today.

If we simply postulate a repulsive force in the universe associated with a
charge density, then we might expect that this force could be responsible
for generating the acceleration of the universe. However, for a
homogeneous and isotropic universe the net charge density would be
zero, although for a finite range force with a small mass
there will exist a non-zero charge density~\cite{Woodard}. The effect of a
Maxwell-type force would be to lower or raise the total energy, leaving the
form of the Friedmann equation unchanged. Thus, we would still have to
invoke exotic forms of energy with an equation of state, $p=w\rho$, where
$w$ is negative and violates the positive energy theorems. For a
non-zero cosmological constant $w=-1$.

In addition to the dark energy problem, we are still confronted
with the puzzle of dark matter. Any observational detection of a
dark matter candidate has eluded us and the fits to galaxy halos
using dark matter models are based on several parameters depending
on the size of the galaxy being fitted. The possibility that an
alternative gravity theory can produce a satisfactory description
of galaxy dynamics with little or no dark matter should be
studied~\cite{Milgrom,Sanders,Mannheim,Aguirre,Bekenstein}.

Challenging experimental results are often the precursors of
a shifting of scientific paradigms. We must now entertain the
prospect that the discovery of the mechanism driving the acceleration of
the universe can profoundly change our description of the universe.
Recent observational data for supernovae (SNe Ia)~\cite{Riess} have
produced new conclusive evidence that the universe went from a decelerating
phase at $z\sim 0.5$ to an accelerating phase and that the data are
consistent with the concordance model with $\Omega_m\sim 0.3,
\Omega_\Lambda\sim 0.7$, where $\Omega_m$ and $\Omega_\Lambda$ denote the
ratios of dark matter density and dark energy density to the critical
density $\rho_c=3H^2/8\pi G$. Moreover, the data give for the equation
of state parameter for dark energy, $w_D=-1.02\pm 0.15$ consistent with
a cosmological constant value $w_D=-1$ and $dw_D/dz= 0$~\cite{Riess}.

Given the uneasy tension existing between observational evidence for the
acceleration of the universe and the mystery of what
constitutes dark matter and dark energy, we shall consider the
question of whether Einstein's general relativity (GR) is correct for the
large scale structure of the universe. It agrees well with local solar
system experimental tests and for the data obtained for observations of
the binary pulsar PSR 1913+16~\cite{Will}. However, this does not preclude
the possibility of a breakdown of the conventional Einstein equations for
the large-scale structure of the universe. The standard GR cosmological
model agrees well with the abundances of light elements from big bang
nucleosynthesis (BBN), and the evolution of the spectrum of primordial
density fluctuations, yielding the observed spectrum of temperature
anisotropies in the cosmic microwave background (CMB). Also, the age of
the universe and the power spectrum of large-scale structure agree
reasonably well with the standard cosmological model. However, it could be
that additional repulsive gravitational effects from an alternative gravity
theory could agree with all of the results in the early universe and yet
lead to significant effects in the present universe accounting for its
acceleration.

A fundamental change in the predictions of the observational data will
presumably only come about from a non-trivial alteration of the
mathematical and geometrical formalism that constitutes GR. Recent
modifications of Einstein's gravitational theory and developments
of cosmological models~\cite{Carroll,Freese,Dvali,Arkani,Kiselev}
have led to alterations of the Friedmann equation at large
cosmological scales.

In the following, we shall consider the
physically non-trivial extension of GR called the nonsymmetric
gravitational theory (NGT)~\cite{Moffat}. This theory was extensively
studied over a period of years, and a version of the theory was discovered
that is free of linearized weak field inconsistencies such as ghost poles,
tachyons, and other instabilities~\cite{Moffat2}. The skew sector
$g_{[\mu\nu]}$ corresponds to a massive Kalb-Ramond field, and is not to
be identified with the electromagnetic field as was done by
Einstein~\cite{Einstein} in his unified field theory. The skew sector is
treated as part of the total gravitational field and together with
the symmetric part $g_{(\mu\nu)}$ forms the total nonsymmetric
$g_{\mu\nu}$, which generates a non-Riemannian geometry.

A significant modification of the
infrared gravitational field and the equations of cosmology can only be
obtained by having additional degrees of freedom beyond the two degrees of
freedom of Einstein's gravity theory. Such additional degrees of freedom
are supplied by NGT, which yields six additional degrees of freedom
in the skew sector $g_{[\mu\nu]}$.

As we shall see in the following, and in further
work~\cite{Moffat3}, it is possible for NGT to describe the
current data on the accelerating universe and the dark matter
halos of galaxies, gravitational lensing and cluster behavior, as
well as the standard observational results, without invoking the
need for dominant, exotic dark matter and a dark energy identified
with vacuum energy and the cosmological constant. These results
are obtained while retaining the good agreement of Einstein's
gravity theory with solar system tests, terrestrial gravitational
experiments and the binary pulsar PSR 1913+16.

\section{NGT Action and Field Equations}

The nonsymmetric $g_{\mu\nu}$ and
$\Gamma^\lambda_{\mu\nu}$ are defined
by~\cite{Moffat,Moffat2}:
\begin{equation}
g_{\mu\nu}=g_{(\mu\nu)}+g_{[\mu\nu]}
\end{equation}
and
\begin{equation}
\Gamma^\lambda_{\mu\nu}=\Gamma^\lambda_{(\mu\nu)}
+\Gamma^\lambda_{[\mu\nu]},
\end{equation}
where
\begin{equation}
g_{(\mu\nu)}={1\over 2}(g_{\mu\nu}+g_{\nu\mu}),\quad g_{[\mu\nu]}= {1\over
2}(g_{\mu\nu}-g_{\nu\mu}).
\end{equation}

The contravariant tensor
$g^{\mu\nu}$ is defined in terms of the equation
\begin{equation}
\label{inverse}
g^{\mu\nu}g_{\sigma\nu}=g^{\nu\mu}g_{\nu\sigma}={\delta^\mu}_\sigma.
\end{equation}

The NGT action is given by
\begin{equation}
S_{\hbox{ngt}}=S+S_M,
\end{equation}
where
\begin{equation}
\label{NGTLagrangian} S=\frac{1}{16\pi G}\int d^4x[{\bf
g}^{\mu\nu}R^*_{\mu\nu}(W)-2\Lambda\sqrt{-g} -{1\over 4}\mu^2{\bf
g}^{\mu\nu}g_{[\nu\mu]}],
\end{equation}
and $S_M$ is the matter action satisfying the relation
\begin{equation}
\frac{1}{\sqrt{-g}}\biggl(\frac{\delta S_M}{\delta
g^{\mu\nu}}\biggr)=-\frac{1}{2}T_{\mu\nu}.
\end{equation}
Here, we have chosen units $c=1$, ${\bf
g}^{\mu\nu}=\sqrt{-g}g^{\mu\nu}$, $g=\hbox{Det}(g_{\mu\nu})$,
$\Lambda$ is the cosmological constant, $\mu$ is a mass associated
with the skew field $g_{[\mu\nu]}$. Moreover, $T_{\mu\nu}$ is the
nonsymmetric energy-momentum tensor and $R^*_{\mu\nu}(W)$ is the
tensor
\begin{equation}
R^*_{\mu\nu}(W)=R_{\mu\nu}(W)-\frac{1}{6}W_\mu W_\nu,
\end{equation}
where $R_{\mu\nu}(W)$ is the NGT contracted curvature tensor
\begin{equation}
R_{\mu\nu}(W)=W^\beta_{\mu\nu,\beta} - {1\over
2}(W^\beta_{\mu\beta,\nu}+W^\beta_{\nu\beta,\mu}) -
W^\beta_{\alpha\nu}W^\alpha_{\mu\beta} +
W^\beta_{\alpha\beta}W^\alpha_{\mu\nu},
\end{equation}
defined in terms of the unconstrained nonsymmetric connection:
\begin{equation}
\label{Wequation}
W^\lambda_{\mu\nu}=\Gamma^\lambda_{\mu\nu}-{2\over
3}{\delta^\lambda}_\mu W_\nu,
\end{equation}
where
\begin{equation}
W_\mu={1\over 2}(W^\lambda_{\mu\lambda}-W^\lambda_{\lambda\mu}).
\end{equation}
Eq.(\ref{Wequation})
leads to the result
\begin{equation}
\Gamma_\mu=\Gamma^\lambda_{[\mu\lambda]}=0.
\end{equation}
The contracted tensor $R_{\mu\nu}(W)$ can be written as
\begin{equation}
R_{\mu\nu}(W)=R_{\mu\nu}(\Gamma)+\frac{2}{3}W_{[\mu,\nu]},
\end{equation}
where
\begin{equation}
R_{\mu\nu}(\Gamma ) = \Gamma^\beta_{\mu\nu,\beta} -{1\over 2}
\left(\Gamma^\beta_{(\mu\beta),\nu} + \Gamma^\beta_{(\nu\beta),\mu}\right)
- \Gamma^\beta_{\alpha\nu} \Gamma^\alpha_{\mu\beta} +
\Gamma^\beta_{(\alpha\beta)}\Gamma^\alpha_{\mu\nu}.
\end{equation}

A variation of the action $S_{\hbox{ngt}}$ yields the field
equations in the presence of matter sources
\begin{equation}
\label{Gequation} G^*_{\mu\nu} (W)+\Lambda g_{\mu\nu}+S_{\mu\nu}
=8\pi GT_{\mu\nu},
\end{equation}
\begin{equation}
\label{divg}
{{\bf g}^{[\mu\nu]}}_{,\nu}=-\frac{1}{2}{\bf g}^{(\mu\alpha)}W_\alpha,
\end{equation}
\begin{equation}
{{\bf g}^{\mu\nu}}_{,\sigma}+{\bf g}^{\rho\nu}W^\mu_{\rho\sigma}
+{\bf g}^{\mu\rho}
W^\nu_{\sigma\rho}-{\bf g}^{\mu\nu}W^\rho_{\sigma\rho}
$$ $$
+{2\over 3}\delta^\nu_\sigma{\bf g}^{\mu\rho}W^\beta_{[\rho\beta]}
+{1\over 6}({\bf g}^{(\mu\beta)}W_\beta\delta^\nu_\sigma
-{\bf g}^{(\nu\beta)}W_\beta\delta^\mu_\sigma)=0.
\end{equation}
Here, we have $G^*_{\mu\nu}(W)=R^*_{\mu\nu}(W) - {1\over 2}
g_{\mu\nu}{\cal R}^*(W)$, where ${\cal
R}^*(W)=g^{\mu\nu}R^*_{\mu\nu}(W)$, and
\begin{equation}
S_{\mu\nu}=\frac{1}{4}\mu^2(g_{[\mu\nu]}
+{1\over 2}g_{\mu\nu}g^{[\sigma\rho]}
g_{[\rho\sigma]}+g^{[\sigma\rho]}g_{\mu\sigma}g_{\rho\nu}).
\end{equation}

The generalized Bianchi identities
\begin{equation}
[{\bf g}^{\alpha\nu}G_{\rho\nu}(\Gamma)+{\bf g}^{\nu\alpha}
G_{\nu\rho}(\Gamma)]_{,\alpha}+{g^{\mu\nu}}_{,\rho}{\bf G}_{\mu\nu}=0,
\end{equation}
give rise to the matter response equations
\begin{equation}
g_{\mu\rho}{{\bf T}^{\mu\nu}}_{,\nu}+g_{\rho\mu}{{\bf T}^{\nu\mu}}_{,\nu}
+(g_{\mu\rho,\nu}+g_{\rho\nu,\mu}-g_{\mu\nu,\rho}){\bf T}^{\mu\nu}=0.
\end{equation}

It has been proved that the present version of NGT described above
does not possess any ghost poles or tachyons in the linear weak
field approximation~\cite{Moffat2}, either as an expansion about
Minkowski spacetime or about a generic GR background. This cures
the inconsistencies discovered by Damour, Deser and McCarthy in an
earlier version of NGT~\cite{Damour}. In NGT there are three
distinct possible metric tensors with three different local light
cone structures. The definition of a spacelike surface is
consequently dependent on the chosen coupling of matter to
geometry and it is not possible to unambiguously apply a $(3+1)$
decomposition of field variables in order to perform a Hamiltonian
constraint analysis using the standard methods. Further studies of
the non-linear and non-perturbative solutions of massive NGT and
its Cauchy development have to be undertaken. In the following, we
shall identify $g_{(\mu\nu)}$ with the metric tensor of spacetime.

\section{Cosmological Solutions}

For the case of a spherically symmetric field, the form of
$g_{\mu\nu}$ in NGT is given by
\begin{equation}
g_{\mu\nu}=\left(\matrix{-\alpha&0&0&w\cr
0&-\beta&f\hbox{sin}\theta&0\cr 0&-f\hbox{sin}\theta&
-\beta\hbox{sin}^2
\theta&0\cr-w&0&0&\gamma\cr}\right),
\end{equation}
where $\alpha,\beta,\gamma$ and $w$ are functions of $r$ and $t$.
The tensor $g^{\mu\nu}$ has the components:
\begin{equation}
g^{\mu\nu}=\left(\matrix{{\gamma\over w^2-
\alpha\gamma}&0&0&{w\over w^2-\alpha\gamma}\cr
0&-{\beta\over \beta^2+f^2}&{f\hbox{csc}\theta\over
\beta^2+f^2}&0\cr
0&-{f\hbox{csc}\theta\over
\beta^2+f^2}&-{\beta\hbox{csc}^2\theta\over
\beta^2+f^2}&0\cr-{w\over w^2-\alpha\gamma}&0&0&-{\alpha\over
w^2-\alpha\gamma}\cr}\right).
\end{equation}

We have
\begin{equation}
\sqrt{-g}=\hbox{sin}\theta[(\alpha\gamma-w^2)(\beta^2+f^2)]^{1/2}.
\end{equation}
For a comoving coordinate system, we obtain for the velocity vector
$u^\mu=dx^\mu/ds$, which satisfies the normalization condition
$g_{(\mu\nu)}u^\mu u^\nu=1$:
\begin{equation}
\label{comovingvelocity}
u^0=\frac{1}{\sqrt{\gamma}},\quad u^r=u^{\theta}=u^{\phi}=0.
\end{equation}

We set $w=0$ so that only the $g_{[23]}$
component of $g_{[\mu\nu]}$ is different from zero. This corresponds to
setting the magnetic monopole charge and static magnetic field in Maxwell's
theory to zero, if we define
\begin{equation}
g^{*[\mu\nu]}=\epsilon^{\mu\nu\sigma\rho}g_{[\sigma\rho]},
\end{equation}
where $\epsilon^{\mu\nu\sigma\rho}$ is the Levi-Civita tensor
density and we identify $g^{*[0i]}$ and $g^{*[jk]}$ as the static
``magnetic'' and ``electric'' potentials, respectively, associated
with the massive Kalb-Ramond potential field.

The vector $W_\mu$ can be determined
from
\begin{equation}
\label{W2equation} W_\mu=-{2\over \sqrt{-g}}s_{(\mu\rho)}{{\bf
g}^{[\rho\sigma]}}_{,\sigma},
\end{equation}
where $s_{(\mu\alpha)}g^{(\alpha\nu)}={\delta^\nu}_\mu$. For the
skew symmetric field with $w=0$, it follows from (\ref{divg}) and
(\ref{W2equation}) that $W_\mu=0$.

The energy-momentum tensor for a fluid is
\begin{equation}
\label{eq:energytensor}
T^{\mu\nu}=(\rho+p)u^\mu u^\nu - pg^{\mu\nu}+K^{[\mu\nu]},
\end{equation}
where $K^{[\mu\nu]}$ is a skew symmetric source tensor identified with the
intrinsic spin or fluid vorticity. The fluid vorticity is defined by
\begin{equation}
\omega_{\mu\nu}=u_{[\mu,\nu]}+a_{[\mu}u_{\nu]},
\end{equation}
where $a_\mu$ is the fluid's four-acceleration.
We only consider
couplings to $u_{[\mu,\nu]}$ to avoid derivative couplings. A rotational
action is given by
\begin{equation}
S_R=\int d^4x{\cal L}_R=\int d^4x\kappa\rho
g_{\mu\nu}\epsilon^{\mu\nu\alpha\beta}u_{[\alpha,\beta]},
\end{equation}
where $\kappa$ is a coupling constant. By varying this action with
respect to $g_{\mu\nu}$ we get
\begin{equation}
K^{[\mu\nu]}=\kappa\rho\frac{\epsilon^{\mu\nu\alpha\beta}}{\sqrt{-g}}u_{[\alpha,\beta]}.
\end{equation}

We define
\begin{equation}
T_{\mu\nu}=g_{\mu\beta}g_{\alpha\nu}T^{\alpha\beta},
\end{equation}
and from (\ref{inverse}) and (\ref{eq:energytensor}), we get
\begin{equation}
T=\rho-3p+g_{[\alpha\beta]}K^{[\alpha\beta]}=\rho - 3p +2fK,
\end{equation}
where $K^{[23]}=K/\sin\theta$.

We can write the field equations (\ref{Gequation}) for $W_\mu=0$ in the
form
\begin{equation}
R_{\mu\nu}(\Gamma)-g_{\mu\nu}\Lambda+S_{\mu\nu}-\frac{1}{2}g_{\mu\nu}{\cal
S} =8\pi G(T_{\mu\nu}-\frac{1}{2}g_{\mu\nu}T),
\end{equation}
where ${\cal S}=g^{\mu\nu}S_{\mu\nu}$. The metric takes the
canonical Gaussian form for comoving polar coordinates
\begin{equation}
ds^2=dt^2-\alpha(r,t)dr^2-\beta(r,t)[d\theta^2+\sin^2\theta d\phi^2].
\end{equation}
We shall assume that $\beta(r,t) \gg f(r,t)$ and that a solution can be found by a
separation of variables
\begin{equation}
\label{separationeq}
\alpha(r,t)=h(r)R^2(t),\quad \beta(r,t)=r^2S^2(t).
\end{equation}

From the field equations (\ref{RSequations})(see Appendix A), we get
\begin{equation}
\label{RSequation}
\frac{{\dot R}}{R}-\frac{{\dot S}}{S}=\frac{1}{2}Zr,
\end{equation}
where ${\dot R}=\partial R/\partial t$ and $Z$ is given by
\begin{equation}
Z=\frac{\dot\beta'f^2} {\beta^3}-\frac{5\dot\beta\beta' f^2}{2\beta^4}
-\frac{\dot\alpha\beta' f^2}{2\alpha\beta^3}+\frac{2\dot\beta ff'}{\beta^3}
-\frac{f\dot f'}{\beta^2}-\frac{3f'\dot f}{2\beta^2}
$$ $$
+\frac{\dot\alpha ff'}{2\alpha\beta^2}+\frac{2\beta'f\dot f}{\beta^3}.
\end{equation}
We shall assume that $Z\approx 0$ which from (\ref{RSequation})
gives $R(t)\approx S(t)$. This leads to a metric of the form
\begin{equation}
\label{FRWmetric}
ds^2=dt^2-R^2(t)\biggl[h(r)dr^2+r^2(d\theta^2+\sin^2\theta
d\phi^2)\biggr].
\end{equation}
We cannot impose exact homogeneity and
isotropy on the skew sector $g_{[\mu\nu]}$, for this would lead
to $f(r,t)=w(r,t)=0$.

We shall further simplify our calculations by assuming that the mass
parameter $\mu \approx 0$,  that the cosmological constant $\Lambda=0$ and
that we can neglect any effects due to the antisymmetric source tensor
$K^{[\mu\nu]}$ associated with vorticity of the matter fluid.

With the assumption that $\beta\gg f$, the equations of motion
become (see, Appendix A):
\begin{equation}
\label{eqn1}
2b(r)+{\ddot{R}}(t)R(t)+2{\dot{R}}^2(t)-R^2(t)W(r,t)=4\pi
GR^2(t)[\rho(r,t)-p(r,t)],
\end{equation}
\begin{equation}
\label{eqn2} -{\ddot R}(t)R(t)+\frac{1}{3}R^2(t)Y(r,t)=\frac{4\pi
G}{3}R^2(t) [\rho(r,t)+3p(r,t)],
\end{equation}
where
\begin{equation}
\label{bequation}
2b(r)={h^\prime(r)\over r h^2(r)}.
\end{equation}
The functions
$W$ and $Y$ are given by
\begin{equation}
\label{Wexpression}
W(r,t)=\frac{\alpha'\beta'f^2}{2\alpha^2\beta^3}
-\frac{\beta^{\prime\prime}f^2}{\alpha\beta^3}
+\frac{\dot\alpha\dot\beta f^2}{2\alpha\beta^3} +\frac{5\beta'^2f^2}
{2\alpha\beta^4}-\frac{\dot\alpha f\dot f}{2\alpha\beta^2}
$$ $$
-\frac{\alpha'ff'}{2\alpha^2\beta^2}-\frac{ff^{\prime\prime}}{\alpha\beta^2}
-\frac{4ff'\beta'}{\alpha\beta^3}+\frac{3f'^2}{2\alpha\beta^2},
\end{equation}
\begin{equation}
\label{Yexpression} Y(r,t)=\frac{\ddot\beta
f^2}{\beta^3}-\frac{5\dot\beta^2f^2}{2\beta^4} -\frac{3\dot
f^2}{2\beta^2}+\frac{4\dot\beta f\dot f}{\beta^3} -\frac{f\ddot
f}{\beta^2}.
\end{equation}
Within our approximation scheme, $W$ and $Y$ can be expressed in the form
\begin{equation}
\label{moreW}
W(r,t)=\frac{h'f^2}{h^2r^5R^6}-\frac{2f^2}{hr^6R^6}+\frac{2{\dot
R}^2f^2}{r^4R^6}+\frac{10f^2}{hr^6R^6}-\frac{{\dot R}f{\dot f}}{r^4R^5}
-\frac{h'ff'}{2h^2r^4R^6}
$$ $$
-\frac{ff''}{h4^4R^6}-\frac{8ff'}{hr^5R^6}
+\frac{3f^{'2}}{2hr^4R^6},
\end{equation}
\begin{equation}
\label{moreY}
Y(r,t)=\frac{2({\dot R}^2+R{\ddot
R})f^2}{r^4R^6}-\frac{10{\dot R}^2f^2}{r^4R^6}-\frac{3{\dot
f}^2}{2r^4R^4}+\frac{8{\dot R}f{\dot f}}{r^4R^5} -\frac{f{\ddot
f}}{r^4R^4}.
\end{equation}

Eliminating ${\ddot R}$ by adding (\ref{eqn1}) and
(\ref{eqn2}), we get
\begin{equation}
\label{Rvelocityeq} \dot{R}^2(t)+b(r)={8\pi G\over 3}\rho(r,t)
R^2(t)+Q(r,t)R^2(t),
\end{equation}
where
\begin{equation}
Q=\frac{1}{2}W-\frac{1}{6}Y.
\end{equation}
From (\ref{eqn2}) we obtain
\begin{equation}
\label{acceleration} {\ddot R(t)}=-\frac{4\pi
G}{3}R(t)[\rho(r,t)+3p(r,t)]+\frac{1}{3}R(t)Y(r,t).
\end{equation}

Let us consider an expansion of $f(r,t)$ about a background
$f_0(r,t)$:
\begin{equation}
f(r,t)=f_0(r,t)+\delta f(r,t)+....
\end{equation}
We shall identify the fluctuations $\delta f(r,t)$ about the
background $f_0(r,t)$ as any matter content additional to
the visible baryon matter of the universe, $\rho_m(r,t)=\rho(\delta
f(r,t))$. Thus, $\rho_m$ replaces the exotic cold dark matter (CDM) of the
standard cosmological model. We also have the expansions
\begin{equation}
W=W_0+\delta W+...;\quad Y=Y_0+\delta Y+....;
$$ $$
Q=Q_0+\delta Q+....
\end{equation}
We define the matter density to be
\begin{equation}
\rho_M=\rho_b+\rho_m,
\end{equation}
where $\rho_b$ denotes the baryon density. The background field $f_0$ will
describe the source of dark energy density and we shall consider slowly
varying solutions of $f_0$, which rise towards the present epoch with red
shift $z\sim 0$ and a solution of $\delta f$ that yields a
$\rho_m=\rho(\delta f)$ that decreases with increasing time as
$1/R^3$.

We can now write in place of (\ref{Rvelocityeq}) and (\ref{acceleration}):
\begin{equation}
\label{Rvelocityeq2} {\dot{R}^2(t)}+b(r)={8\pi G\over
3}\rho_M(r,t)R^2(t)+Q_0(r,t)R^2(t), \end{equation} and
\begin{equation} \label{acceleration2} {\ddot
R}(t)=-\frac{4\pi
G}{3}R(t)[\rho_M(r,t)+3p_M(r,t)]+\frac{1}{3}R(t)Y_0(r,t),
\end{equation} where $Q_0(r,t)=Q(f_0(r,t))$ and
$Y_0(r,t)=Y(f_0(r,t))$.

We can write Eq.(\ref{Rvelocityeq2}) as
\begin{equation}
H^2+\frac{b}{R^2}=\Omega H^2,
\end{equation}
where $H={\dot R}/R$,
\begin{equation}
\Omega=\Omega_M+\Omega_{f_0},
\end{equation}
and
\begin{equation}
\Omega_M=\frac{8\pi G\rho_M}{3H^2},\quad
\Omega_{f_0}=\frac{Q_0}{H^2}.
\end{equation}
From Eq.(\ref{bequation}) for $h=1$, we obtain $b=0$ and $\Omega=1$
and
\begin{equation}
\label{FlatFried} H^2=\frac{8\pi G}{3}\rho_M+Q_0,
\end{equation}
which describes a spatially flat universe. The line element takes the
approximate form of a flat, homogeneous and isotropic FRW universe
\begin{equation}
ds^2=dt^2-R^2(t)[dr^2+r^2(d\theta^2+\sin^2\theta d\phi^2)].
\end{equation}

\section{Accelerating Expansion of the Universe}

Let us assume that there exists a solution of the field equations such that
asymptotically for large $r$ and $t$:
\begin{equation}
Q_0^\prime(r,t)\sim 0,\quad Y_0^\prime(r,t)\sim 0,
\end{equation}
and
\begin{equation}
\rho_M(r,t)\rightarrow \rho_M(t),\quad p_M(r,t)\rightarrow p_M(t).
\end{equation}
Here, we choose the big bang to begin on a hypersurface $\Sigma(r,t)$ with
$r=t=0$, so that asymptotically for large $r$ and $t$ we approach the
present universe.

It follows from (\ref{acceleration2}) that ${\ddot R} > 0$ when $Y_0
> 4\pi G(\rho_M+3p_M)$. If we assume that there exist solutions for
$Q_0(r,t)$ and $Y_0(r,t)$, such that they are sufficiently small in the
early universe, then we will retain the good agreement of Einstein gravity
with the BBN era with $\rho_{\rm rad}\propto 1/R^4$. As the universe
expands beyond the BBN era at the temperatures, $T\sim 60$ kev-1 MeV, and
$\rho_M\propto 1/R^3$, then we must seek solutions such that
$Q_0$ and $Y_0$ begin to increase and reach slowly varying values with
$\Omega^0_{f_0}\sim 0.7$ and $\Omega^0_M\sim 0.3$, where $\Omega^0_M$ and
$\Omega^0_{f_0}$ denote the present values of $\Omega_M$ and
$\Omega_{f_0}$, respectively. Provided we can find solutions of the field
equations that satisfy these conditions, then it should be possible to fit
the combined supernovae, cluster and CMB data~\cite{Riess}.

We observe from (\ref{moreW}) and (\ref{moreY}) that $Q_0$ and
$Y_0$ are functions of the behavior of $R$ and $f$ and their
derivatives. If $f_0$ grows sufficiently with $R$ as $t$
increases, then $Q_0$ and $Y_0$ can dominate the matter
contribution $\rho_M$ as the universe evolves towards the current
epoch with $p_M\approx 0$. A detailed solution of the field
equations is required to determine the dynamical behavior of $R$,
$f$, $Q$ and $Y$.

In the present epoch, $p_M\approx 0$ and (\ref{acceleration2})
gives
\begin{equation}
\frac{\ddot R}{R}=-\frac{1}{3}[4\pi G\rho_M-Y_0].
\end{equation}
We can now define an effective equation of state parameter for the
universe:
\begin{equation}
w_{\rm eff}=\frac{1}{3}(4\pi G\rho_M-Y_0).
\end{equation}
We see that $Y_0 > 4\pi G\rho_M$ corresponds to the usual
condition for acceleration $w_D < -1/3$ with the dark energy
equation of state $p_D=w_D\rho_D$.

We can explain the evolution of Hubble expansion acceleration
within NGT, without violating the positive energy conditions for
matter and radiation. We satisfy the strong energy condition for
matter $\rho_M+3p_M >0$ throughout the evolution of the universe
and {\it there is no need for a cosmological constant}. The $Q_0$
and $Y_0$ contributions to the expansion of the universe increase
at a slow rate up to values today with $Y_0 >4\pi G\rho_M$
($p_M\approx 0$). The cosmological constant $\Lambda=0$ during
this evolution.

We must also guarantee that the influence of the $Y$ and $Q$ contributions
do not conflict with galaxy formation, i.e. the additional NGT
contributions to the Friedmann equation must not couple too strongly with
the attractive gravitational effects predicted by the field equations in
the galaxy formation epoch. This issue and the other required evolutionary
effects of $Q$ and $Y$ must be determined by a numerical computation of
the NGT field equations.

\section{Conclusions}

We have proposed that NGT may explain, as gravitational phenomena,
the accelerating universe without unknown dark energy, and the
observed flat rotation curves of galaxies without the undetected
exotic dark matter~\cite{Moffat3}. We do expect that there is some
dark matter in the universe in the form of dark baryons and
neutrinos with non-vanishing mass. Such a theory can be falsified
with data, whereas it is difficult to falsify the dark energy and
dark matter hypotheses.

NGT would be required to explain the formation
of galaxy structure without CDM. We know that the standard $\Lambda$CDM
model of structure formation and the description of the CMB power spectrum
is remarkably successful, so that solutions of the NGT equations must
produce an equally successful description of the data. To investigate in
detail whether NGT can succeed in producing a successful account of the
large-scale structure of the universe will require solving the field
equations. These are issues that require further investigation.

\section{Appendix A: The $\Gamma$-Connections and Field Equations}

The NGT compatibility equation is given by
\begin{equation}
g_{\lambda\nu,\eta}-g_{\rho\nu}\Gamma^\rho_{\lambda\eta}
-g_{\lambda\rho}\Gamma^\rho_{\eta\nu}=\frac{1}{6}g^{(\mu\rho)}
(g_{\rho\nu}g_{\lambda\eta}-g_{\eta\nu}g_{\lambda\rho}-
g_{\lambda\nu}g_{[\rho\eta]})W_\mu,
\end{equation}
where $W_\mu$ is determined from (\ref{W2equation}). When
$w(r,t)=g_{[01]}(r,t)=0$, it follows that $W_\mu=0$ and the compatibility
equation reads
\begin{equation}
g_{\lambda\nu,\eta}-g_{\rho\nu}\Gamma^\rho_{\lambda\eta}
-g_{\lambda\rho}\Gamma^\rho_{\eta\nu}=0.
\end{equation}

The non-vanishing components of the $\Gamma$-connections are:
\begin{eqnarray}
\Gamma^1_{11}&=&\frac{\alpha'}{2\alpha},\\
\Gamma^1_{(10)}&=&\frac{\dot{\alpha}}{2\alpha},\\
\Gamma^1_{22}&=&\Gamma^1_{33}\hbox{cosec}^2\theta
=\frac{1}{2\alpha}\biggl(fB-\frac{1}{2}\beta A'\biggr),\\
\Gamma^1_{00}&=&\frac{\gamma'}{2\alpha},\\
\Gamma^2_{(12)}&=&\Gamma^3_{(13)}=\frac{1}{4}A',\\
\Gamma^2_{(20)}&=&\Gamma^3_{(30)}=\frac{1}{4}\dot{A},\\
\Gamma^2_{33}&=&-\sin\theta\cos\theta,\\
\Gamma^3_{(23)}&=&\cot\theta,\\
\Gamma^0_{(11)}&=&\frac{\dot{\alpha}}{2\gamma},\\
\Gamma^0_{(10)}&=&\frac{\gamma'}{2\gamma},\\
\Gamma^0_{22}&=&\Gamma^0_{33}\hbox{cosec}^2\theta
=- \frac{1}{2\gamma}\biggl(fD-\frac{1}{2}\beta\dot{A}\biggr),\\
\Gamma^0_{00}&=&\frac{\dot{\gamma}}{2\gamma},\\
\Gamma^1_{[23]}&=&\frac{\sin\theta}{2\alpha}\biggl(\frac{1}{2}fA'+\beta
B\biggl),\\
\Gamma^2_{[13]}&=&-\Gamma^3_{[12]}\sin^2\theta=\frac{1}{2}B\sin\theta,\\
\Gamma^2_{[30]}&=&- \Gamma^3_{[20]}\sin^2\theta=-
\frac{1}{2}D\sin\theta,\\
\Gamma^0_{[23]}&=&-
\frac{\sin\theta}{2\gamma}\biggl(\frac{1}{2}f\dot{A}
+\beta D\biggl),
\end{eqnarray}
where $A, B$ and $D$ are given by
\begin{equation}
A=\ln(\beta^2+f^2),
\end{equation}
\begin{equation}
B=\frac{f\beta^\prime-\beta f^\prime}{\beta^2+f^2},
\end{equation}
and
\begin{equation}
D=\frac{{\dot\beta}f-{\dot f}\beta}{\beta^2+f^2}.
\end{equation}

The NGT field equations in the presence of sources are given by
\begin{equation}
R_{11}(\Gamma)=-{1\over 2}A^{''}-{1\over 8}[(A^\prime)^2+4B^2]
+{\alpha^\prime A^\prime\over 4\alpha}
+{\gamma^\prime\over 2\gamma}\biggl({\alpha^\prime\over 2\alpha}
-{\gamma^\prime\over 2\gamma}\biggr)
$$ $$
-\biggl({\gamma^\prime\over 2\gamma}\biggr)^\prime
+{\partial\over \partial t}\biggl({{\dot \alpha}\over 2\gamma}\biggr)
+{{\dot \alpha}\over 2\gamma}\biggl({{\dot\gamma}\over 2\gamma}
-{{\dot\alpha}\over 2\alpha}
+{1\over 2}{\dot A}\biggr)
+\Lambda\alpha-\frac{1}{4}\mu^2\frac{\alpha
f^2}{\beta^2+f^2}
$$ $$
=4\pi G\alpha(\rho-p+2fK),
\end{equation}
\begin{equation}
R_{22}(\Gamma)=
R_{33}(\Gamma)\hbox{cosec}^2\theta=1+ \biggl({2fB-\beta A^\prime\over
4\alpha}\biggr)^\prime +\biggl({2fB-\beta A^\prime\over
8\alpha^2\gamma}\biggr) (\alpha^\prime\gamma+\gamma^\prime\alpha)
$$ $$
+{B(fA^\prime+2\beta B)\over 4\alpha}
-{\partial\over \partial t}\biggl({2fD-\beta{\dot A}\over
4\gamma}\biggr)
-{2fD-\beta{\dot A}\over 8\alpha\gamma^2}({\dot\alpha}\gamma
+{\dot \gamma}{\alpha})
$$ $$
-{D\over 4\gamma}(f{\dot A}+2\beta
D)+\Lambda\beta+\frac{1}{4}\mu^2\frac{\beta f^2}{\beta^2+f^2}
=4\pi G\beta(\rho-p +2fK),
\end{equation}
\begin{equation}
R_{00}(\Gamma)=-{1\over 2}{\ddot A}-{1\over
8}({\dot A}^2+4D^2) +{{\dot\gamma}\over 4\gamma}{\dot
A}+{{\dot\alpha}\over 2\alpha}\biggl({{\dot\gamma}\over 2\gamma}
-{{\dot\alpha}\over 2\alpha}\biggr)
$$ $$
-{\partial\over \partial t}\biggl({{\dot\alpha}\over 2\alpha}\biggr)
+\biggl({\gamma^\prime\over 2\alpha}\biggr)^\prime
+{\gamma^\prime\over 2\alpha}\biggl({\alpha^\prime\over 2\alpha}
-{\gamma^\prime\over 2\gamma}
+{1\over 2}A^\prime\biggr)
$$ $$
-\Lambda\gamma+\frac{1}{4}\mu^2\frac{\gamma f^2}{\beta^2+f^2}
=4\pi G\gamma(\rho+3p - 2fK),
\end{equation}
\begin{equation}
\label{[10]equation} R_{[10]}(\Gamma)=8\pi GK_{[10]}=0,
\end{equation}
\begin{equation}
\label{RSequations}
R_{(10)}(\Gamma)= -\frac{1}{2}{\dot A}'+\frac{1}{4}A'
\biggl(\frac{{\dot\alpha}}{\alpha}-\frac{1}{2}{\dot A}\biggr)
+\frac{1}{4}\frac{\gamma' {\dot A}}{\gamma}-\frac{1}{2}BD=0,
\end{equation}
\begin{equation}
R_{[23]}(\Gamma)=\sin\theta\biggl[\biggl({fA^\prime+2\beta
B\over 4\alpha}\biggr)^\prime+{1\over 8\alpha}(fA^\prime+2\beta B)
\biggl({\alpha^\prime\over \alpha}+{\gamma^\prime\over\gamma}\biggr)
$$ $$
-{B\over 4\alpha}(2fB-\beta A^\prime)-{1\over
8\gamma}(f{\dot A}+2\beta D) \biggl({{\dot\gamma}\over
\gamma}+{{\dot\alpha}\over \alpha}\biggr)
$$ $$
-{\partial\over \partial t}\biggl({f{\dot A}+2\beta D\over
4\gamma}\biggr)
+{D\over 4\gamma}(2fD-\beta{\dot A})\biggr]
$$ $$
-\biggl[\Lambda f-\frac{1}{4}\mu^2
f\biggl(1+\frac{\beta^2}{\beta^2+f^2}\biggr)\biggr]\sin\theta= -
4\pi Gf\sin\theta(\rho - p).
\end{equation}

\vskip 0.2 true in
{\bf Acknowledgments}
\vskip 0.2 true in

This work was supported by the Natural Sciences and Engineering Research
Council of Canada. I thank Robert Brandenberger, Michael Clayton
and Martin Green for helpful and stimulating discussions.
\vskip
0.5 true in

 \end{document}